\documentclass{article}
\usepackage[a4paper, margin=1in]{geometry}

\usepackage{amsfonts}
\usepackage{amsmath}
\usepackage{multicol}
\usepackage{booktabs}
\usepackage{multirow}
\usepackage{algorithm,algpseudocode}
\usepackage{siunitx}
\usepackage{stfloats}
\usepackage{footmisc}

\usepackage{tikz}
\usepackage[european]{circuitikz}
\usetikzlibrary{positioning}
\usetikzlibrary{fit}
\usetikzlibrary{backgrounds}
\usetikzlibrary{calc}
\usetikzlibrary{shapes}
\usetikzlibrary{shadows}
\usetikzlibrary{chains}
\usetikzlibrary{trees}
\usetikzlibrary{arrows}
\usetikzlibrary{intersections}
\usetikzlibrary{mindmap}
\usetikzlibrary{decorations.text}
\usetikzlibrary{spy}
\usetikzlibrary{patterns}
\usetikzlibrary{shapes.geometric, arrows}

\newcommand{\footremember}[2]{%
    \footnote{#2}
    \newcounter{#1}
    \setcounter{#1}{\value{footnote}}%
}
\newcommand{\footrecall}[1]{%
    \footnotemark[\value{#1}]%
} 

\begin{document}

\title{Identification of Power System Dynamic Model Parameters using the Fisher Information Matrix}

\author{
Dawn Virginillo,\footremember{epfl}{D. Virginillo and M. Paolone are with the Distributed Electrical Systems Laboratory, \'{E}cole Polytechnique F\'{e}d\'{e}rale de Lausanne, Lausanne, Switzerland (e-mail: \{dawn.virginillo; mario.paolone\}@epfl.ch).}\footremember{corr}{Corresponding author.} Asja Dervi\v{s}kadi\'{c},\footremember{sg}{A. Dervi\v{s}kadi\'{c} is with Swissgrid AG, Aarau, Switzerland (e-mail: asja.derviskadic@swissgrid.ch).} and Mario Paolone\footrecall{epfl}
}

\maketitle

\begin{abstract}
The expected decrease in system inertia and frequency stability motivates the development and maintenance of dynamic system models by Transmission System Operators.  However, some dynamic model parameters can be unavailable due to market unbundling, or inaccurate due to aging infrastructure, non-documented tuning of controllers, or other factors.  In this paper, we propose the use of a numerical approximation of the Fisher Information Matrix (nFIM) for efficient inference of dynamic model parameters. Thanks to the proposed numerical implementation, the method is scalable to Electromagnetic Transient (EMT) models, which can quickly become computationally complex even for small study systems.  Case studies show that the nFIM is coherent with parameter variances of single- and multi-parameter least-squares estimators when applied to an IEEE 9-bus dynamic model with artificial measurements.

\end{abstract}

\section{Introduction}\label{Intro}

There is a constant need for Transmission System Operators (TSOs) to develop and maintain accurate dynamic models to study electromechanical transients.  Well-tuned dynamic models can be used to assess potential problems associated with the operation of the grid and also to define detailed operational sequences in preparation for emergency situations (i.e., grid restoration after a blackout).  However, due to market unbundling, not all dynamic model parameters are always available to TSOs~\cite{Feltes2003,Coughlan2007}. Further, it has been shown that dynamic models of power systems can suffer from inadequate validation due to outdated parameters, in the case of aging infrastructure~\cite{Gonzales2020,Podlaski2022}.  Inadequate documentation of synchronous machine controller tuning processes, or a lack of coordination between involved parties, can also result in models which are improperly calibrated.

The focus of this work is to investigate quantitative strategies for dynamic modelling of power plants in the absence of detailed models, or where the available models are not fit-for-purpose to the intended application.  In this paper, we propose the use of the Fisher Information Matrix (FIM)~\cite{kay}, and specifically, a numerical approximation of the same, to enable efficient dynamic model validation using test measurements.  

To this end, we use the case study of synchronous machine controller parameters, as synchronous machines still form the backbone of the current electrical network and support the integration of DERs through the provision of ancillary services; however, these strategies could also be deployed for other power plant types or for parameters of other components in a dynamic model.  We demonstrate the application of the proposed method using a modified IEEE 9-bus standard transmission network model~\cite{pscontrolstab}.  To the authors' knowledge, this is the first application of Fisher information theory for the estimation of synchronous machine and controller parameters.

\noindent The contributions of this paper are summarized as follows:
\begin{itemize}
     \item Proposal of a new method for quantitatively informed and efficient power system dynamic modelling in the absence of detailed controller knowledge, and 
     \item Demonstration of how the numerical Fisher Information Matrix can be used to identify parameters of synchronous machine controllers.
\end{itemize}

The paper is organized as follows: in Section~\ref{sec:lit}, we present the state of the art on power system identification methods, focusing on dynamic system studies.  In Section~\ref{sec:fim}, we present the use of FIM for the identification of parameters of dynamic power systems. In Section~\ref{sec:cstudy}, we show application examples which demonstrate the proposed method. Finally, we draw conclusions in Section~\ref{sec:conc}.

\section{State of the Art}\label{sec:lit}

Within the power systems domain, but also in the literature dealing with system identification, a variety of parameter estimation techniques have been proposed.  These include conventional least-squares methods, Kalman filter-based methods, and data-driven techniques.  This review gives examples in each category, and concludes with a survey on existing applications of Fisher information theory to system identification problems.  

A generic identification problem is described as follows (adapted from~\cite{Podlaski2022}).  Given a set of observed measurements $z(t)$, we aim to find the set of parameters $p$ which results in a simulated system response which follows the measurements as closely as possible.  If the simulated system response is denoted as $y(t,p)$, a generic optimization problem can be formulated as follows:

\begin{subequations}\label{eq:generic_id}
    \begin{alignat}{5}
        & \min_{p} & \textrm{ } & || z(t) - y(t,p) ||_k  &  \label{eq:of} \\
        &\text{s.t.} & &  y(t,p) = f(t,p,x,u) \label{eq:constr}
    \end{alignat}
\end{subequations}

\noindent where the objective function of Problem~(\ref{eq:generic_id}), i.e., Eq.~(\ref{eq:of}), minimizes the $k$-norm of the difference between the measurements and simulated response, and the constraints of Problem~(\ref{eq:generic_id}), i.e., Eq.~(\ref{eq:constr}), specifies that the simulated system response is a function of time $t$, parameters $p$, states $x$, and the input $u$. The function $f$, representing the mathematical form of the simulated response, is a multi-dimensional set of algebraic and differential equations. Problem~(\ref{eq:generic_id}) is generally non-convex and solved using gradient descent techniques~\cite{id-stoica,kay}. It is initialized using a white-box model of the system with assumed parameters $p$, which can be based on hypotheses or prior knowledge.  In practice, the measurements $z(t)$ can be collected either in real-time operation or during pre-installation manufacturer tests.

Least-squares estimators are commonly applied to identification problems by minimizing the squared error between a simulated output and available measurements (i.e., the squared 2-norm implementation of Problem~(\ref{eq:generic_id})).  For example, in~\cite{Fabbiani2022}, the authors use a least-square method to estimate the network AC admittance matrix.  
This method has also been used to estimate machine mechanical parameters using modal analysis data~\cite{Guo2014,Gorbunov2020}.
In addition,~\cite{Lokhov2018} uses PMU measurements to estimate the state of the system for model validation using least squares estimators.  This technique has also been applied to estimate parameters of dynamic models, as in~\cite{Podlaski2022}, and specifically for the purpose of restoration, for example, in~\cite{Paolone-book}.  The main drawback of least-squares estimation is that due to non-linearity, convergence to a global optimum is not guaranteed~\cite{Conte2020}.

Kalman filters and their variations have been applied in the power systems literature for purposes of both State Estimation (SE) and parameter estimation problems.  For example, in~\cite{Gonzales2020}, the authors estimate the parameters of a wind turbine setup, including a synchronous generator and voltage source converter, using cubature Kalman filters.  Extended Kalman filters are applied in~\cite{Pereira2022} for the estimation of line parameters using measurement data.  The authors of~\cite{Ghahremani2011} apply unscented Kalman filters for estimation of synchronous machine states.  
It is worth noting that the convergence of a Kalman filter to the ground truth is guaranteed if both the measurement and process models are linear and the corresponding measurement and process covariance noise models are white.  Therefore, the extended Kalman filter cannot guarantee convergence, and further, cannot guarantee convergence to the ground truth~\cite{id-dynamic-book}.

Various data-driven and machine learning techniques have been explored for the estimation of power system parameters.  Machine learning techniques provide advantages due to the non-convexity of Problem~(\ref{eq:generic_id}) and the simulation time required for multiple evaluations of the simulated response $y(t,p)$~\cite{Stiasny2024}.  For example, in~\cite{Kontis2017}, neural networks are tested for the identification of dynamic load parameters.  The authors of~\cite{ahmed2006} propose a genetic algorithm for the identification of Automatic Voltage Regulator (AVR) parameters.  Particle Swarm Optimization is proposed for a similar problem in~\cite{Cheng2011}.  However, it is known that these methods struggle to handle events outside of those contained in the training data~\cite{Stiasny2024}, and there is no guarantee of a unique solution due to the non-convex formulation.

The Fisher Information Matrix (FIM) is a technique which has been well-studied in the context of signal processing~\cite{Pakrooh2013,id-stoica}, and is commonly deployed in the literature for quantitatively-informed design of experiments~\cite{Sovljanski2024} (see Section~\ref{sec:fim}).  In~\cite{Pakrooh2013}, the authors derive the analytical Cramér-Rao Lower Bound (CRLB) for parameter estimation given compressed measurement data.  The authors of~\cite{Niazadeh2012} provide commentary on the achievability of the CRLB by an estimator under conditions of non-Gaussian measurement noise, demonstrating that CRLB can also be applied to nonstandard problem conditions.  

There is increasing interest in the literature on the application of FIM to the field of dynamic power system modeling.  In~\cite{Fabbiani2022}, the voltage setpoints in a distribution grid are modified according to Fisher information theory to estimate the network admittance matrix using a least-squares estimator.  Some further examples are~\cite{Xygkis2018}, in which Fisher information theory is used in a PMU placement algorithm designed to maximize the accuracy of state estimation, and~\cite{Halihal2024}, in which an analytical CRLB is derived for an estimator of the system admittance matrix.  However, to the authors' knowledge, Fisher information theory has not yet been demonstrated in the context of power system dynamics for the estimation of synchronous machine, governor, and AVR parameters. For this field of application, the constraints of Problem~(\ref{eq:generic_id}) can be assumed to contain differential equations, complexifying the estimation problem and providing a use case for the numerical FIM.

\section{On the Use of the Fisher Information Matrix}\label{sec:fim}

\subsection{Definition of the Hypotheses}
A generic identification problem is described as follows~\cite{kay,bock2007}.  Suppose that a dynamical system is modelled with output $y = f(t,u,x,\tilde{p})$, which is dependent on the time $t$, input $u$, states $x$, and a `true' parameter set $\tilde{p}$. The sensor with which the generic output $z_i$ is measured at time $t_i$ has uncertainty $\epsilon_i$, according to the following relation:

\begin{equation}
    z_i = y(t_i,u,x,\tilde{p}) + \epsilon_i
\end{equation}

The goal of an identification problem is to estimate the set of parameters $\hat{p}$ which corresponds to the real set of parameters $\tilde{p}$ with maximum likelihood.  The estimator can then be characterized by the following generic equation:

\begin{equation}
    \hat{p} = T(z)
\end{equation}
\noindent where $T$ refers to the estimator, and $z$ refers to the observed system output (i.e., the set of collected measurements with uncertainty).  Here, we suppose that the estimator $T$ is unbiased, namely: the expected value $\mathbb{E}$ of the set of estimated parameters is equal to the set of true parameters.  This can be expressed mathematically as follows:

\begin{equation}
    \mathbb{E} (\hat{p}) - \tilde{p} = 0
\end{equation}

Further, we suppose that the estimator $T$ is consistent, meaning that the estimated parameters $\hat{p}$ converge to a mean bias value $\bar{p} = \mathbb{E} (\hat{p}) - \tilde{p}$ if the total number of observations $N$ is infinitely many.  This can be expressed mathematically as follows:

\begin{equation}
    \lim_{N \rightarrow \infty} T(z_1,z_2,...,z_N) = \bar{p}
\end{equation}

It should be noted that, for an unbiased estimator, $\bar{p} = 0$.  The Cramér-Rao Lower Bound (CRLB), which is defined as the minimum of the reciprocal of the Fisher Information~\cite{kay,id-stoica,lin2005}, defines a lower limit on the variance of an unbiased estimator $T$ which gives the supposed set of parameters $\hat{p}$ as an output.  The definition of the CRLB can be mathematically expressed for a single parameter $p_1$ as follows:

\begin{equation}
    \mathrm{var}(T(z)) \geq \frac{1}{\mathcal{I}(z)}
\end{equation}

\noindent where $\mathcal{I}(z)$ is the scalar Fisher information carried in a set of observations $z$.  This can also be expressed for multiple parameters in matrix form (i.e., using the covariance of the estimator $\mathbf{T}$) as follows:

\begin{equation}
    \mathrm{cov}(\mathbf{T}(z)) \geq {\mathbf{I}(z)}^{-1}
\end{equation}

\noindent where $\mathbf{I}(z)$ is called the Fisher Information Matrix (FIM).  The CRLB can be used to anticipate the performance of estimators using different measurements for the same parameter set by virtue of the expected output parameter variances, as is commonly applied in the field of optimal experimental design~\cite{Sovljanski2024,kay}.  As for our case, the constraints of Problem~(\ref{eq:generic_id}) are a set of algebraic and differential equations, in this paper, we estimate the Fisher information of different measurements \textit{numerically} and apply the measurements with the highest information content to achieve the best-fit parameters with highest confidence.

\subsection{Derivation of the Numerical Fisher Information Matrix}
In this section, a numerical approximation of the FIM is derived. The FIM is defined as the second derivative of the log-likelihood function~\cite{kay}.

\underline{\textit{Fundamental Observation 1}}: As anticipated, the application domain of interest of this paper is a power system consisting of synchronous machines, transformers, lines, and loads, making the constraints of Problem~(\ref{eq:generic_id}) a set of algebraic and differential equations.  Suppose that we are interested in a voltage transient in a power system as measured by the RMS voltage at the stator terminals of a single machine.  In theory, it is possible to derive the log-likelihood score function, or Fisher information with respect to a given parameter (i.e., the gain of the AVR), using the above equation, with the variable $y(t_i,u,x,p)$ defined as the simulated RMS voltage with a certain set of parameters $p$.  However, in this case, the derivative of the voltage with respect to the parameter is also needed.  In the case of the synchronous machine, this would involve many additional variables, as the terminal voltage depends on the flux and current in the d- and q-axes, with limited practical advantages to obtaining an analytical solution.  Here, it is proposed to use the nFIM to inform the design of the final estimator $T$.

It is further noted that in this case, as synchronous machine controllers have many parameters, the Fisher information should be expressed in matrix form.  It is assumed that the applied power system model defines the relation between the input $u$ and output $y$ of the system, such that the uncertainty in the measurements can be characterized by white Gaussian noise. This hypothesis appears reasonable due to the types of uncertainties characterizing current and potential transformers (CTs, PTs) commonly used in power systems~\cite{std-ct,std-vt}.

First, if the measurement noise is assumed to be Gaussian and unbiased (i.e., the mean of the noise is zero), the Probability Distribution Function (PDF) of the measurements for a single sample $f_i(z_i,p)$ is:
\begin{equation}
    f_i(z_i,p) = \frac{1}{\sqrt{2 \pi \sigma^2}}e^{-\frac{(z_i-y(t_i,u,x,p))^2}{2 \sigma_n^2}}
\end{equation}

\noindent where $\sigma_n^2$ is the variance of the measurement noise and $y(t_i,u,x,p)$ is the simulated output with a given set of parameters $p$ and for a given timestep $t_i$.  For a set of measurements $z$, the PDF of the measurements $f(z,p)$ is defined as:
\begin{equation}
    f(z,p) = \frac{1}{\sqrt{2 \pi \sigma^2}^N}e^{-\frac{\sum_i(z_i-y(t_i,u,x,p))^2}{2 \sigma_n^2}}
\end{equation}

\noindent where $N$ is the total number of observations.  Further, the log-likelihood function is defined as the logarithm of the probability distribution function for the full set of measurements~\cite{kay}.  Thus:
\begin{equation}
    \ln(f(z,p)) = -\ln \left[ \sqrt{2 \pi \sigma^2}^N \right] - \frac{1}{2 \sigma_n^2} \sum_i(z_i-y(t_i,u,x,p))^2
\end{equation}

The log-likelihood score function (Fisher information with respect to parameter $p_k$) can be expressed analytically as:
\begin{equation}\label{eq:fim}
    \begin{split}
    \mathcal{I}(p_k) &= \frac{\partial \ln(f(z,p))}{\partial p_k} \\
                     &= -\frac{1}{\sigma_n^2} \sum_i(z_i-y(t_i,u,x,p)) \frac{\partial y(t_i,u,x,p)}{\partial p_k}
    \end{split}
\end{equation}

To compute the nFIM, we employ the following numerical approximation for a given parameter $p_k$:
\begin{equation}
    \frac{\partial \ln(f(z,p))}{\partial p_k} \approx \frac{\ln(f(z,(1+\alpha_k)p_k))-\ln(f(z,p_k))}{\alpha_k p_k}
\end{equation}

\noindent where $\alpha_k$ is a perturbation applied to the starting parameter $p_k$.  For ease of notation, we define $p_{\alpha k} = (1+\alpha_k) p_k$.  It is noted that in this case, the simulated output $y$ also depends on the value of the parameter $p_k$.  By substitution, we arrive at the following expression:
\begin{equation}\label{eq:approx}
    \frac{\partial \ln(f(z,p))}{\partial p_k} \approx \frac{\sum_i(z_i-y(t_i,p_{\alpha k}))^2 - \sum_i(z_i-y(t_i,p_{k}))^2}{2\sigma_n^2 \alpha_k p_k}
\end{equation}

To put the nFIM in matrix form, we observe the analytical formulation for the $m,k$th element of FIM, which is defined as follows:
\begin{equation}
    \mathbf{I}_{m, k}(p) = \mathbb{E}\left[\frac{\partial}{\partial p_m} \ln (f(z, p)) \cdot \frac{\partial}{\partial p_k} \ln (f(z, p))\right]
\end{equation}

Thus, the diagonal elements are defined as follows:
\begin{equation}
    n\mathbf{I}_{k,k}(p) = \left[\frac{\sum_i(z_i-y(t_i,p_{\alpha k}))^2 - \sum_i(z_i-y(t_i,p_{k}))^2}{2\sigma_n^2 \alpha_k p_k} \right ]^2
\end{equation}

Further, the off-diagonal elements are defined as follows:
\begin{equation}
\begin{split}
    n&\mathbf{I}_{k,j}(p) = \frac{1}{4 \sigma_n^4} \\
    & \cdot \left[\frac{\sum_i(z_i-y(t_i,p_j,p_{\alpha k}))^2 - \sum_i(z_i-y(t_i,p_j,p_{k}))^2}{\alpha_k p_k} \right] \\
        & \cdot \left[\frac{\sum_i(z_i-y(t_i,p_{\alpha j},p_k))^2 - \sum_i(z_i-y(t_i,p_j,p_{k}))^2}{\alpha_j p_j} \right]
\end{split}
\end{equation}

It is noted that the perturbation applied to parameter $j$ is not necessarily equal to that applied to parameter $k$, that is to say that for certain cases, $\alpha_j \neq \alpha_k$.

\underline{\textit{Fundamental Observation 2}}: the nFIM represents a characterization of the information present in a set of noisy measurements which uses a numerical approximation of the transient $y$ and which does not depend on the partial derivative of the transient $y$ with respect to parameter $p_k$.  As such, the nFIM enables the use of a numerical simulation model of the power grid without the need to know the analytical formulation of the `true' transient $y(t_i,u,x,\tilde{p})$.

It can be seen that, in this formulation, the choice of the perturbation $\alpha_k$ and the initial value of the parameter $p_k$ have an impact on the estimation of the nFIM.  To maximize the accuracy of the nFIM estimate, the starting parameter $p_k$ should be as close as possible to the true parameter $\tilde{p}$.  In practice, the optimized parameter value(s) at the end of a least squares optimization can be used, as the true parameter $\tilde{p}$ is unknown.  Further, the perturbation $\alpha_k$ should result in a change in the simulated transient which is enough to overcome the measurement noise.  To quantify this, we define the following relation:

\begin{equation}\label{eq:stdev}
    \sigma_d > C \sigma_n
\end{equation}

\noindent where $\sigma_d$ is the standard deviation of the difference between the two simulated curves $d(\alpha_k) = y(t_i,p_j,p_{\alpha k}) - y(t_i,p_j,p_{k})$, $\sigma_n$ is the standard deviation of the noise, and $C$ is a constant which is greater or equal to 1.

To apply these relations, the noise variance $\sigma_n^2$ should be computed according to the following relation for a certain desired signal-to-noise ratio:

\begin{equation}
    \sigma^2 = \left[ \frac{\frac{1}{N} \sum_i z_i}{10^{\textrm{SNR}/20}} \right]^2
\end{equation}

In words, the variance of the noise is equal to the squared mean of the noisy signal (measurements) divided by the squared signal-to-noise ratio amplitude.

To capture the effects of both diagonal and off-diagonal elements, we use the volume of the ellipsoid spanned by the FIM to assess the confidence region~\cite{Sovljanski2024}.  For a larger ellipsoid volume, a higher nCRLB, and thus an estimator with higher variances in the parameters, can be expected.  The confidence ellipsoid volume is defined as follows:

\begin{equation}\label{eq:ve}
    V_e = \frac{2 \pi^{P/2}}{P \Gamma(P/2)} \prod_i \sqrt{1/\lambda_i}
\end{equation}

\noindent where $P$ is the number of parameters considered (equivalent to the dimension of FIM), $\Gamma$ is the gamma function, and $\lambda$ are the eigenvalues of FIM.  This is commonly referred to as E-optimal experimental design~\cite{Sovljanski2024}.

\subsection{Nonlinear Optimization of Parameters for Identification}
In the absence of a detailed synchronous machine controller model, it is necessary to validate dynamic models using measurements.  As the problem is nonlinear, an iterative least-squares approach can be applied to determine the most likely true controller parameters given a set of measurements.  The main drawback of such methods is that convergence to a global minimum cannot be guaranteed~\cite{conv-opt}. 
Therefore, it is proposed to apply a least-squares estimator on a subset of parameters $p_\lambda$. 

The mathematical formulation of the optimization problem is as follows, for a generic system with PQ-loads at buses $B_L$ and synchronous machines at buses $B_M$.  It is noted that the formulation is an adaptation of Problem~(\ref{eq:generic_id}).
\newcommand*{\bfrac}[2]{\genfrac{[}{]}{0pt}{}{#1}{#2}}
\begin{subequations}
    \begin{alignat}{3}
        & \min_{p_\lambda} & \quad & \sum_i \left [ z_i - y_i(t_i,p_\lambda) \right]^2  \\
        &\text{s.t.} & &  \mathbf{i} = G\mathbf{v}    \label{eq:subb} \\
        &  & &  i_b = -\frac{v_b^2}{L_b} \quad \forall b \in B_L \label{eq:subd}\\
        &  & &  \Psi_b = MI_b \quad \forall b \in B_M \label{eq:sube}\\
        &  & &  V_b = -RI_b + \frac{d\Psi_b}{dt} + \bfrac{ -\omega_e \Psi_{q,b}}{\omega_e \Psi_{d,b}} \quad \forall b \in B_M  \label{eq:subf} \\
        &  & &  J_b\frac{d\omega_{m,b}}{dt} + D_b\theta_b = \Delta T_b \quad \forall b \in B_M  \label{eq:subg} \\
        &  & &  P_{m,b} = T_b \omega_{m,b} \quad \forall b \in B_M  \label{eq:subg2} \\
        &  & &  P_{m,b} = G_b(p_\lambda) \cdot \Delta \omega_e \quad \forall b \in B_M \label{eq:subh} \\
        &  & &  E_{fd,b} = H_b(p_\lambda) \cdot V_{b,\textrm{RMS}} \quad \forall b \in B_M \label{eq:subi} \\
        &  & & \omega_e = \frac{1}{2} n_{p,1} \omega_{m,1}  \label{eq:subi2} \\
        &  & & p_l \leq p_\lambda \leq p_u  \label{eq:subj} 
    \end{alignat}
\end{subequations}

\noindent where $z$ is a set of measurements, and $y(t,p)$ is the time-dependent simulated output of the selected quantity (i.e., mechanical or electrical frequency, active or reactive power, etc.) generated with parameter set $p$. Constraint~\ref{eq:subb} represents the relation between voltages and currents at each node, where $G$ is the time-dependent nodal admittance matrix of the system, and $\mathbf{v}$ and $\mathbf{i}$ are the nodal voltage and current injection vectors, respectively~\cite{psanalysis}.  Here, the nodal admittance matrix $Y$ contains the parameters of the synchronous machine step-up transformers and high-voltage transmission lines.  Constraint~\ref{eq:subd} computes the elements of the nodal current injection vector for the load buses, given the user-defined complex power $L_b$, which can also be time-dependent. 

The synchronous machines are characterized by Constraints~\ref{eq:sube}--\ref{eq:subi}~\cite{jean2011,std-sync}.  Constraint~\ref{eq:sube} relates the magnetic flux on each winding of the synchronous machine to its current through a matrix of inductances $M$, which is defined by the provided synchronous machine parameters.  Constraint~\ref{eq:subf} relates the winding voltages to currents and fluxes through a matrix of winding resistances R and the electrical frequency $\omega_e$.  $\Psi_{d,b}$ and $\Psi_{q,b}$ represent the magnetic flux on the d- and q-axes of the synchronous machine at bus $b$, respectively.  Constraint~\ref{eq:subg} models the mechanical reaction of the synchronous machine rotating masses, where $\omega_m$ represents the mechanical speed, $J$ the moments of inertia of the rotating masses, $D$ the user-defined angular damping constants, and $\Delta T$ is the difference in torque between the turbine and generator.  Constraint~\ref{eq:subg2} defines the relation between mechanical power and torque.  The control actions of the governor and AVR are modelled by equations~\ref{eq:subh} and~\ref{eq:subi}, respectively, where $G_b$ and $H_b$ are the transfer functions of the governor and AVR.  $P_m$ represents the mechanical power applied to the rotor, $E_{fd}$ represents the voltage applied to the rotor winding, and $V_{b,\textrm{rms}}$ represents the Root-Means-Squared (RMS) voltage at bus $b$.  

The electrical frequency is defined according to the mechanical speed of generator 1 in Constraint~\ref{eq:subi2}.  Finally, Constraint~\ref{eq:subj} defines upper and lower bound vectors for the parameter set $p_\lambda$.

This problem is made nonlinear due to constraints~\ref{eq:subd}, ~\ref{eq:subf}, ~\ref{eq:subg}, ~\ref{eq:subh}, and~\ref{eq:subi}.  In practice, as per Fundamental Observation 2, this can be implemented using an iterative least squares solver connected with software which models power system dynamics, as illustrated by the following formulation:

\begin{subequations}
    \begin{alignat}{5}
        & \min_{p_\lambda} & \textrm{ } & \sum_i \left [ z_i - y_i(t_i,p_\lambda) \right]^2 &   \\
        &\text{s.t.} & &  
            \textrm{solution for } y \textrm{ using a dynamic power} \\  
            &&&\textrm{system modelling software} \nonumber \\
        &                  & & p_l \leq p_\lambda \leq p_u & 
    \end{alignat}
\end{subequations}

It is noted that the subset of parameters which is not included in the optimization set, $p_\mu \notin p_\lambda$, are not validated using this initial approach.  Depending on the available computational power and time for validation, the method can be applied to optimize parameters $p_\mu$ in an iterative fashion to achieve the best possible result.  However, the advantage of this approach is that it allows a modeler to perform model validation in a structured, quantitatively informed, and efficient manner.

\section{Application Examples}\label{sec:cstudy}
In this section, example cases which validate the usefulness and functionality of the method are given using a dynamic version of the IEEE 9-bus model~\cite{pscontrolstab}.  First, the details of the 9-bus model are given.  Then, a case study with single-parameter estimators is used to demonstrate the coherency of the nFIM implementation.  Finally, a more realistic case study, in which multiple parameters are to be estimated, is shown.

The following procedure is proposed to identify the parameters of a power island, as shown in Algorithm~\ref{alg:1}.  First, the system of interest is modeled in detail using a set of base, or default, parameters, and selected initial transients are generated.  Then, the numerical Fisher Information Matrix (nFIM) is used to identify the set of measurements which carry the highest information about the parameter(s) of interest.  As discussed in Section~\ref{sec:fim} and shown in Algorithm~\ref{alg:1}, the measurements are compared using the confidence ellipsoid volume spanned by nFIM, which is computed using Eq.~\ref{eq:stdev},\ref{eq:ve}.  The set of measurements $z$ exhibiting the smallest confidence ellipsoid volume $V_e$ is chosen.  Finally, a nonlinear gradient descent optimization is applied to optimize the selected parameters $p_\lambda$ such that the dynamic model achieves the best least-squares fit relative to the measurements.  This approach enables the exploitation of known characteristics of the dynamic power system (physics-based models), in contrast with black-box methods, and is scalable to large systems with many unknown parameters.  Here, we focus on the application of the algorithm to identifying synchronous machine controller parameters.

\begin{algorithm}
\caption{Dynamic model validation using nFIM}\label{alg:1}
\begin{algorithmic}
\State \textbf{Initialize:} $p_\lambda = p_0$
\For{$z \in Z$}
    \State Compute or assume $\sigma_n$
    \State Choose $C$ to satisfy Eq. (\ref{eq:stdev})
    \State Compute $V_e$ using Eq. (\ref{eq:ve})
\EndFor
\State $z = \texttt{argmin}(V_e)$
\While{$\lvert \Delta \sum_i \left [z_i-y_i(t_i,p_\lambda)\right ]^2 \lvert /\sum_i \left [z_i-y_i(t_i,p_\lambda) \right]^2 >\zeta$ $\And \lVert D \cdot (p_{\lambda+1}-p_\lambda) \lVert_2 > \eta$}
    \State $p_\lambda=H(p_{\lambda-1})$
\EndWhile
\end{algorithmic}
\end{algorithm}

For the purpose of method validation, we apply Algorithm~\ref{alg:1} for a certain number $N$ of artificial  measurement sets, such that the output contains $N$ sets of parameters $p_\lambda$, and the variances of the parameters can be compared to the average volume of the confidence ellipsoid spanned by nFIM $V_e$. 

The deployed computational environment is described as follows.  The IEEE 9-bus model is implemented in EMTP~\cite{EMTP}.  Matlab is used for the FIM computations~\cite{MATLAB:2023}, and the simulation results are imported into MATLAB from EMTP using~\cite{EMTP-to-matlab}.  The least squares estimator described in~\cite{electrimacs} is used for the parameter identification with the Trust Region Reflective algorithm~\cite{matlab-opt}.  The Trust Region Reflective algorithm defines the parameter update function $H$ during the least squares optimization, and is configured to stop when either the function convergence threshold $\zeta$ or the step size convergence threshold $\eta$ is reached~\cite{matlab-lsqnonlin}.\footnote{In this implementation, both convergence thresholds $\zeta$ and $\eta$ are 1E-6.}

\subsection{Dynamic Modelling of the IEEE 9-bus System}
\begin{figure}[ht]
    \centering
    \resizebox{9cm}{4.85cm}{
    \begin{tikzpicture}[x=3cm,y=3cm]
        \coordinate                    (BUS2) at (0.5,0);
        \coordinate                    (BUS7) at (1,0);
        \coordinate                    (BUS8) at (2,0);
        \coordinate                    (BUS9) at (3,0);
        \coordinate                    (BUS3) at (3.5,0);
        \coordinate                    (BUS5) at (1.5,-0.5);
        \coordinate                    (BUS6) at (2.5,-0.5);
        \coordinate                    (BUS4) at (2,-1);
        \coordinate                    (BUS1) at (2,-1.5);
        \coordinate                    (BUS7_2) at (1,-0.1);
        \coordinate                    (BUS8_2) at (2,-0.1);
        \coordinate                    (BUS9_2) at (3,-0.1);
        \coordinate                    (BUS4_2) at (1.9,-1);
        \coordinate                    (BUS4_3) at (2.1,-1);
        \coordinate                    (BUS5_2) at (1.4,-0.5);
        \coordinate                    (BUS5_3) at (1.6,-0.5);
        \coordinate                    (BUS6_2) at (2.6,-0.5);
        \coordinate                    (BUS6_3) at (2.4,-0.5);
        \coordinate                    (GEN1) at (2,-2);
        \coordinate                    (GEN2) at (0,0);
        \coordinate                    (GEN3) at (4,0);
    
        \draw 
            (BUS7) to[short,*-*]                (BUS8)
                   to[short,*-*]                (BUS9);
        \draw (BUS7_2) to[short,*-] (1.4,-0.1) to[short,-*] (BUS5_2);
        \draw (BUS9_2) to[short,*-] (2.6,-0.1) to[short,-*] (BUS6_2);
        \draw (BUS5_3) to[short,*-] (1.6,-0.6) -- (1.9,-0.9) to[short,-*] (BUS4_2);
        \draw (BUS6_3) to[short,*-] (2.4,-0.6) -- (2.1,-0.9) to[short,-*] (BUS4_3);
        
        \draw [thick] (0.5,-0.15) -- (0.5,0.15) node[above]{Bus 2};
        \draw [thick] (1,-0.15) -- (1,0.15) node[above]{Bus 7};
        \draw [thick] (2,-0.15) -- (2,0.15) node[above]{Bus 8};
        \draw [thick] (3,-0.15) -- (3,0.15) node[above]{Bus 9};
        \draw [thick] (3.5,-0.15) -- (3.5,0.15) node[above]{Bus 3};
        \draw [thick] (1.35,-0.5) node[left]{Bus 5} -- (1.65,-0.5);
        \draw [thick] (2.35,-0.5) -- (2.65,-0.5) node[right]{Bus 6};
        \draw [thick] (1.85,-1) -- (2.15,-1) node[right]{Bus 4};
        \draw [thick] (1.85,-1.5) -- (2.15,-1.5) node[right]{Bus 1};
        
        \draw (GEN2) node[vsourcesinshape, rotate=90, anchor=north](S){} (S.south) to[short,-*] (BUS2);
        \draw (GEN3) node[vsourcesinshape, rotate=90, anchor=south](S){} (S.north) to[short,-*] (BUS3);
        \draw (GEN1) node[vsourcesinshape, rotate=90, anchor=west](S){} (S.east) to[short,-*] (BUS1);
        \draw (GEN1) node[below]{$SM_1$};
        \draw (GEN2) node[left]{$SM_2$};
        \draw (GEN3) node[right]{$SM_3$};

        \draw (BUS2) to [oosourcetrans] (BUS7);
        \draw (BUS9) to [oosourcetrans] (BUS3);
        \draw (BUS4) to [oosourcetrans,*-] (BUS1);
        
        \draw [-stealth](BUS5) to[short,*-] (1.5,-0.7) node[left]{$L_5$};
        \draw [-stealth](BUS6) to[short,*-] (2.5,-0.7) node[right]{$L_6$};
        \draw [-stealth](BUS8_2) to[short,*-] (2.1,-0.1) -- (2.1,-0.3) node[right]{$L_8$};
    \end{tikzpicture}
    }
    \caption{IEEE 9-bus model~\cite{pscontrolstab}.}
    \label{fig:9busmodel}
\end{figure}
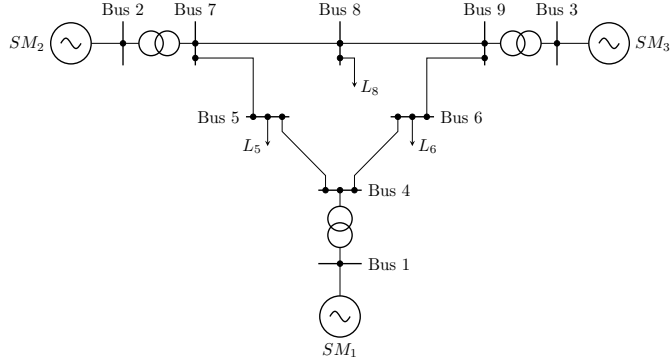

To demonstrate the advantages of using the nFIM for dynamic power system model identification, the IEEE 9-bus model is used.  The dynamic simulations are initialized from the results of the load flow simulation given in Table~\ref{tab:lf-9bus}, adapted from~\cite{pscontrolstab}. Saturation is not modeled, but could be added in case relevant phenomena are to be studied.  A single-line diagram of the study system is shown in Figure~\ref{fig:9busmodel}. 

\begin{table}[ht]
\caption{Load and Generator Data for 9-Bus Model\cite{pscontrolstab}}
\centering
\begin{tabular}{c|c|c|c}
\toprule
Machine & \begin{tabular}[c]{@{}c@{}}Nominal \\ Voltage (kV) \end{tabular} & \begin{tabular}[c]{@{}c@{}}Active \\ Power (MW) \end{tabular} & \begin{tabular}[c]{@{}c@{}}Reactive \\ Power (Mvar) \end{tabular} \\ \midrule
$SM_1$  & 16.5   & 71.6 & 27    \\ \midrule   
$SM_2$  & 18     & 163  & 6.7   \\ \midrule  
$SM_3$  & 13.8   & 85   & -10.9 \\ \midrule  
$L_5$   & 230    & 125  & 50    \\ \midrule 
$L_6$   & 230    & 90   & 30    \\ \midrule  
$L_8$   & 230    & 100  & 35    \\ \bottomrule
\end{tabular}
\label{tab:lf-9bus}
\end{table}

The standard EMTP models are used for the 2-winding transformers~\cite{EMTP-trafo} and transmission lines~\cite{EMTP-line}.  The load models are constant RLC. The generators are modelled using the EMTP synchronous machine model~\cite{EMTP-SM} with standard parameters.

For both case studies, the energization of a $15 \textrm{~}\si{MW}$ load at Bus 6 is the modelled operation.  Therefore, the original configuration with $L_6 = 90 \textrm{~}\si{MW}$~\cite{pscontrolstab} is modified such that $75 \textrm{~}\si{MW}$ is initially energized, followed by an additional $15 \textrm{~}\si{MW}$.

For the purposes of the study, the structure of the synchronous machine controllers is assumed as a precondition using standard models.  The Automatic Voltage Regulator (AVR) is of type SEXS~\cite{spd-avr}, and the governor is of type IEEEG3~\cite{turb-gov-argonne}. Power System Stabilizers (PSS) are neglected.  The block diagrams of the governor and AVR are shown in Figures~\ref{fig:gov} and~\ref{fig:avr}, respectively.  The initialization points $P_\mathrm{ref}$ and $V_\mathrm{ref}$ are computed according to the following relations:

\begin{equation}
    P_\mathrm{ref} = \frac{K_\sigma}{K_t}\times P_\mathrm{mss}
\end{equation}
\begin{equation}
    V_\mathrm{ref} = \frac{1}{K_\mathrm{AVR}}\times E_{fd\mathrm{ss}} + V^0_\mathrm{rms}
\end{equation}

\noindent where $P_\mathrm{mss}$ and $E_{fd\mathrm{ss}}$ are the steady-state mechanical power and field voltage, respectively, and $V^0_\mathrm{rms}$ is the initial RMS voltage.

\tikzset{
block/.style = {draw, fill=white, rectangle, minimum height=3em, minimum width=3em},
tmp/.style  = {coordinate}, 
sum/.style= {draw, fill=white, circle, node distance=1cm},
input/.style = {coordinate},
output/.style= {coordinate},
pinstyle/.style = {pin edge={to-,thin,black}}
}

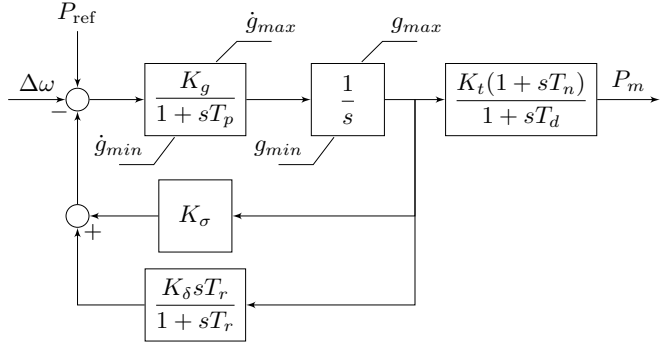
\begin{figure}[ht]
\centering
\resizebox{8.8cm}{4.6cm}{
\begin{tikzpicture}[auto, node distance=2cm,>=latex']
    \node [input, name=rinput] (rinput) {};
    \node [sum, right of=rinput] (sum1) {};
    \node [block, right of=sum1, node distance=1.7cm] (servo) {$\dfrac{K_g}{1+sT_p}$};
    \node [block, right of=sum1, node distance=3.9cm] (int) {$\dfrac{1}{s}$};
    \node [block, right of=int,node distance=2.5cm] (turb) {$\dfrac{K_t (1+sT_n)}{1+sT_d}$};
    \node [block, below of=servo,node distance=1.7cm] (sigma) {$K_\sigma$};
    \node [block, below of=sigma,node distance=1.3cm] (delta) {$\dfrac{K_\delta sT_r}{1+sT_r}$};
    \node [sum, below of=sum1,node distance=1.7cm] (sum2) {};
    \node [output, right of=turb, node distance=2cm] (output) {};
    \draw [->] (rinput) -- node{$\Delta \omega$} (sum1);
    \draw [->] (sum1) -- (servo);
    \draw [->] (servo) -- (int);
    \draw [->] (int) --node[name=z,anchor=south]{} (turb);
    \draw [->] (sigma) -- node[pos=0.95] {$+$} (sum2);
    \draw [->] (turb) -- node [name=y] {$P_m$}(output);
    \draw [->] (z) |- (delta);
    \draw [->] (z) |- (sigma);
    \draw [->] (delta) -| (sum2);
    \draw [->] (sum2) -- node[pos=0.99] {$-$} (sum1);
    \draw [->] ($(0,1.0cm)+(sum1)$)node[above]{$P_{\mathrm{ref}}$} -- (sum1);
    \draw (3,0.55) -- (3.3,0.85) -- (3.8,0.85) node[above]{$\dot{g}_{max}$};
    \draw (5.1,0.55) -- (5.4,0.85) -- (5.9,0.85) node[above]{${g}_{max}$};
    \draw (2.4,-0.55) -- (2.1,-0.95) -- (1.6,-0.95) node[above]{$\dot{g}_{min}$};
    \draw (4.7,-0.55) -- (4.4,-0.95) -- (3.9,-0.95) node[above]{${g}_{min}$};
    \end{tikzpicture}
    }
\caption{Block diagram of implemented governor/turbine model IEEEG3~\cite{turb-gov-argonne}.} \label{fig:gov}
\end{figure}

\begin{figure}[ht]
\centering
\begin{tikzpicture}[auto, node distance=2cm,>=latex']
    \node [input, name=rinput] (rinput) {};
    \node [sum, right of=rinput, node distance=1.2cm] (sum1) {};
    \node [block, right of=sum1, node distance=1.5cm] (avr1) {$\dfrac{1+sT_a}{1+sT_b}$};
    \node [block, right of=avr1, node distance=2.7cm] (avr2) {$\dfrac{K}{1+sT_e}$};
    \node [output, right of=avr2, node distance=2cm] (output) {};
    \draw [->] (rinput) -- node{$V_\mathrm{rms}$} (sum1);
    \draw [->] (sum1) -- (avr1);
    \draw [->] (avr1) -- (avr2);
    \draw [->] (avr2) -- node [name=y] {$E_{fd}$}(output);
    \draw ($(-0.3cm,0)+(sum1)$)node[below]{$-$};
    \draw [->] ($(0,1.0cm)+(sum1)$)node[above]{$V_{\mathrm{ref}}$} -- (sum1);
    \draw (5.7,0.55) -- (6,0.85) -- (6.6,0.85) node[above]{${E}_{fd,max}$};
    \draw (5.1,-0.55) -- (4.8,-0.95) -- (4.2,-0.95) node[above]{$E_{fd,min}$};
    \end{tikzpicture}
\caption{Block diagram of implemented AVR model SEXS~\cite{spd-avr}.} \label{fig:avr}
\end{figure}
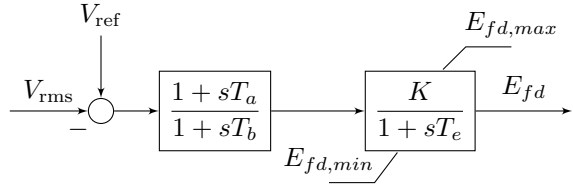

To create the `true' parameters $\tilde{p}$, the controller parameters are first tuned to produce a stable response, adapted from the parameter set in~\cite{Kundur}.  In the absence of detailed controller models, it is assumed that a modeler would not have access to a simulation with the true parameters $\tilde{p}$.  Therefore, a random perturbation between -10\% and +10\% of the true parameter value is applied to each parameter, using a random value generated using a uniform distribution, resulting in a set of `Initial Parameters' $p_0$.  The initial parameter set $p_0$ and true parameter set $\tilde{p}$ for synchronous machine $SM_1$ are given in Table~\ref{tab:params}.  Only transfer function Time Constants (TC), coefficients (CF), and gains are listed.  

Although a description of only the relevant model information is provided here, with some omissions for brevity, the full model has been published and is available for the interested reader at~\cite{github-9bus}.

\begin{table}
\caption{True and initial parameter sets for $SM_1$.}
\centering
\begin{tabular}{c|c|c}
\toprule
Parameter & $\tilde{p}$ & $p_0$ \\ \midrule
AVR Numerator TC $T_a$ [s]          & 1        & 1.0526 \\ \midrule   
AVR Denominator TC $T_b$ [s]        & 12       & 11.3898 \\ \midrule  
Exciter TC $T_e$ [s]                & 0.04     & 0.0401 \\ \midrule  
Exciter gain $K$ [p.u.]             & 20       & 18.5881 \\ \midrule 
Gate gain $K_g$ [p.u.]              & 5        & 4.6297 \\ \midrule  
Valve TC $T_p$ [s]                  & 0.05     & 0.0516 \\ \midrule  
Transient droop TC $T_r$ [s]        & 5        & 4.5732 \\ \midrule   
Permanent droop CF $\sigma$ [p.u.]  & 0.04     & 0.0376  \\ \midrule
Transient droop CF $\delta$ [p.u.]  & 0.8      & 0.8561 \\ \midrule
Turbine gain $K_t$ [p.u.]           & 1.5      & 1.5287 \\ \midrule
Turbine numerator TC $T_n$ [s]      & -1.7067  & -1.7672 \\ \midrule
Turbine denominator TC $T_d$ [s]    & 2.4      & 2.6232 \\ \bottomrule
\end{tabular}
\label{tab:params}
\end{table}

\subsection{Case Study: Single-Parameter Estimators}

A single-parameter estimator study is presented to demonstrate the accuracy of the nFIM implementation, and the usefulness of the nFIM method.  Here, we compare the results of Algorithm~\ref{alg:1} using the same set of measurements with five different target parameters.  The objective is to show that the parameter variances are coherent with the estimated nFIM.

\subsubsection{Description of the Measurements}
For this study, the frequency of the system is examined using the measured rotational speed of the synchronous machine at bus 1, $\omega_{m1}$.  The examined transient is shown in the lower half of Figure~\ref{fig:meas}.  The measurements are created by adding artificial Gaussian white noise to a simulation of the transient $\omega_{m1}$ with the true parameters $\tilde{p}$, which are given in Table~\ref{tab:params}.  For this case study, a Signal-to-Noise Ratio (SNR) of 80 dB is used.  

The noise variance, needed for the formulation of nFIM, is computed using the following relation:

\begin{equation}\label{eq:snr}
    \sigma_n^2 = \left [ \frac{\frac{1}{N}\sum_{n=1}^N y(t_n,p_0)}{10^{\textrm{SNR}/20}} \right]^2
\end{equation}

\noindent where $y(t_i,\tilde{p})$ is the simulated frequency with the true set of parameters.   It is further noted that the noise variance is in units of the transient, which is, in this case, radians per second (rad/s).  Therefore, for this case, $\sigma_n = 0.0063 \textrm{~}\si{rad/s}$.  In practice, the noise variance could be inferred from the specifications of a real measurement device. 

\subsubsection{Application of Algorithm~\ref{alg:1}}
For this case study, the five parameters of interest are the AVR gain $K$, the permanent droop coefficient $\sigma$, the transient droop coefficient $\delta$, the turbine gain $K_t$, and the turbine denominator time constant $T_d$. Each result has a single parameter in the set $p_\lambda$ and therefore a scalar value for the Fisher information (no off-diagonal elements).  The parameters which are not in the set to be optimized $p_{\lambda i}$ for a given optimization run $i$ are set to their true values $\tilde{p}$.  

The set of `Initial Parameters' $p_0$, created by applying perturbations on the true parameter values, is given in Table~\ref{tab:params}.  In practice, the initial parameters should be an initial guess, either from a priori knowledge about the controllers (i.e., from the manufacturer or a previously developed model), or from the literature (i.e., from~\cite{Kundur}).

As the work uses a \textit{numerical} approximation of the FIM, it is prudent to comment on how the numerical implementation impacts the approximation.  First, it can be said that so long as the least-squares optimization algorithm converges towards the true value as measured by the residual norm, a better approximation is achieved at the final point than at the initial point.  Therefore, for this case study, we consider only the nFIM at the final optimized value to ensure a fair comparison between the estimators for different parameters.

\begin{figure}[h]
    \centering
    \includegraphics[width=8.89cm]{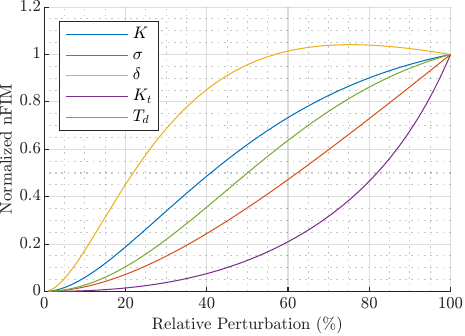}
    \caption{Impact of perturbation $\alpha$ on the nFIM for the five considered parameters using a single set of measurements.}
    \label{fig:pert}
\end{figure}

Further, the perturbation applied to compute the nFIM $\alpha_k$ also impacts the approximation.  We observe this effect by varying the applied perturbation relative to the parameter value (i.e., in \% of the optimized parameter value $p_\lambda$).  For this purpose, a single set of measurements is used (`Set 1'), and the results for the five parameters are shown in Figure~\ref{fig:pert}.  Specifically, the nFIM normalized by the nFIM with a 100\% relative perturbation is shown for each parameter.  It is observed that the nFIM increases initially steeply, then stabilizes as the relative perturbation increases, but that this stabilization point occurs at different relative perturbation values for each parameter.

From the results, it can thus be deduced that there is a trade-off between the accuracy of the numerical derivative approximation and having a perturbation which is significant enough to overcome the noise in the measurements.  For large relative perturbations, the accuracy of Eq.~(\ref{eq:approx}) diminishes.  For the purposes of comparison of estimators for different parameters, for this case study, we use Eq.~(\ref{eq:stdev}) and select $C$ close to 1.

\subsubsection{Results}

The results of the single-parameter estimator analysis are shown in Table~\ref{tab:single-param-res} with $N = 100$ simulations (i.e., 100 unique sets of artificial measurements generated from the same simulated transient).  The same set of measurements is used for each of the 5 estimators for different parameters.  The shown nFIM and $V_e$ figures are averaged across the 100 simulation runs and computed with the final optimized parameter value.  As the standard deviation of the noise is $0.0063 \textrm{~}\si{rad/s}$, the relative perturbation of each parameter is chosen such that $\sigma_d$ is close to, but greater than, this value.  The standard deviation of the parameters $\sigma_p$ is converted to per-unit values for appropriate comparison.

Most notably, it can be seen that the ranking of the variance of the parameters $\sigma^2_p$ matches the ranking of the confidence ellipsoid volume.  This confirms the hypothesis that the relative confidence ellipsoid volumes between estimators for different parameters can be used to anticipate the resulting parameter variances.

Further, as a verification, the parameter variances can be compared to the nCRLB, which is the reciprocal of the scalar Fisher information.  The nCRLB, as a theoretical lower bound on the estimator performance, should always be smaller than the achieved parameter variance.  The results satisfy this criterion for all estimators, demonstrating the applicability of Algorithm~\ref{alg:1}.

\begin{table*}[ht]
  \centering
  \caption{Results of the Single-Parameter Estimator Study.}
  \resizebox{\columnwidth}{!}{%
  \begin{tabular}{c|c|c|c|c|c|c|c|c|c|c|c}
    \toprule
    Parameter & $\tilde{p}$ & $\alpha$ & \begin{tabular}[c]{@{}c@{}}Pert. \\ Abs. \end{tabular} & $\sigma_d$ & \begin{tabular}[c]{@{}c@{}}Avg. \\ nFIM \end{tabular} & \begin{tabular}[c]{@{}c@{}}Avg. \\ nCRLB \end{tabular} & \begin{tabular}[c]{@{}c@{}}Avg. \\ $V_e$ \end{tabular} & Rank & \begin{tabular}[c]{@{}c@{}}Avg.\\Relative \\ Error (\%) \end{tabular} & $\sigma^2_p$ [p.u.] & Rank \\ \midrule
    $K$      & 20   & 0.15   & 3       & 0.0068  & 1.27E11 & 7.87E-12 & 5.61E-6 & 4 & 0.00012 & 4.58E-6  & 4 \\ \midrule
    $\sigma$ & 0.04 & 0.425  & 0.017   & 0.0065  & 9.04E15 & 1.11E-16 & 2.10E-8 & 1 & 0.0033  & 7.84E-11 & 1 \\ \midrule
    $\delta$ & 0.8  & 0.0275 & 0.022   & 0.0068  & 2.27E15 & 4.41E-16 & 4.20E-8 & 2 & 0.00024 & 2.17E-10 & 2 \\ \midrule
    $K_t$    & 1.5  & 0.0225 & 0.03375 & 0.00634 & 7.51E14 & 1.33E-15 & 7.29E-8 & 3 & 0.00015 & 1.15E-9  & 3 \\ \midrule
    $T_d$    & 2.4  & 0.04   & 0.096   & 0.0064  & 3.70E10 & 2.70E-11 & 1.04E-5 & 5 & 0.00004 & 1.63E-5  & 5 \\ \bottomrule
  \end{tabular}
  }
  \label{tab:single-param-res}
\end{table*}

\subsection{Case Study: Multi-Parameter Estimators}

It is noted that in a realistic dynamic modelling scenario, a certain parameter set of interest is defined for the tuning and validation process, and more than one parameter is unknown.  According to the method in Algorithm~\ref{alg:1}, in this paper, we propose the use of nFIM to quantitatively select which transient should be used to tune the parameter set of interest $p_\lambda$.  In this multi-parameter estimator study, two different measurement sets are considered, with all 5 parameters in the set $p_\lambda$ initialized to their assigned values $p_0$.  The average confidence ellipsoid volume $\bar{V_e}$ across the chosen number of simulations $N$ is then compared to the average parameter variance $\bar{\sigma}_p^2$  to validate the functionality of Algorithm~\ref{alg:1}.

\subsubsection{Description of the Measurements}

\begin{figure}[t]
    \centering
    \includegraphics[width=8.89cm]{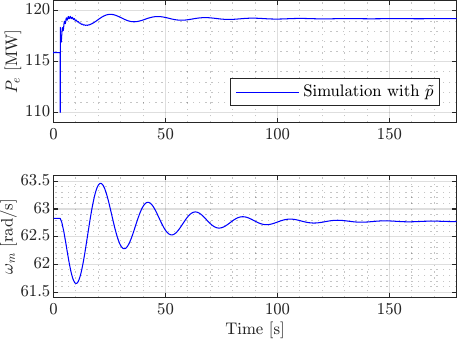}
    \caption{Simulation of $SM_1$ active power $P_e$ (above) and rotational speed $\omega_m$ (below) with true parameter set $\tilde{p}$.}
    \label{fig:meas}
\end{figure}

For the multi-parameter estimator study, each of the two measurement sets is generated by adding artificial white Gaussian noise to the respective simulated quantity.  Given that the operation of interest is the energization of a purely active power load, frequency and active power transients are selected for examination.  The frequency measurements are the same as those used for the single-parameter estimator study, specifically, the mechanical rotational speed of synchronous machine 1 ($\omega_m$ for $SM_1$).  The active power measurements are generated using the electrical power $P_e$ of the synchronous machine 1.  The two selected transients are shown in Figure~\ref{fig:meas}.

Due to the difference in magnitude between the two signals, the standard deviation of the noise $\sigma_n$ are $0.0063 \textrm{~}\si{rad/s}$ and $0.0119 \textrm{~}\si{MW}$ for the frequency and power transients, respectively, computed using Eq.~(\ref{eq:snr}).  80 dB of artificial noise is added for both cases.

\subsubsection{Application of Algorithm~\ref{alg:1}}

As for the single-parameter estimator study, the relative perturbation $\alpha_k$ for each of the 5 parameters is chosen such that $\sigma_d$ is slightly above the noise variance $\sigma_n$.  For this study, the number of experiments $N=100$.

\begin{figure}[t]
    \centering
    \includegraphics[width=8.89cm]{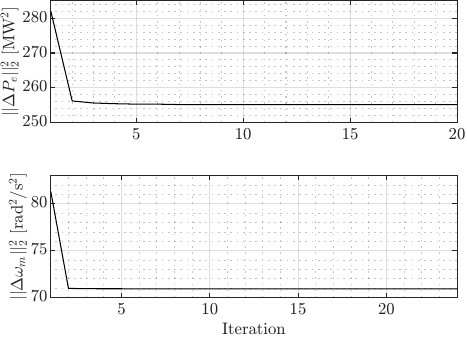}
    \caption{Convergence of the least-squares optimizer with measurement Set 1 of each of the two considered transients $P_e$ (above) and $\omega_m$ (below).}
    \label{fig:conv}
\end{figure}

As Algorithm~\ref{alg:1} includes a least-squares optimization, it is also prudent to comment on the convergence of the optimization problem.  The convergence of the estimator for the first set of generated measurements `Set 1' for each of the two studied transients is shown in Figure~\ref{fig:conv} in terms of the squared 2-norm of the error between the measurements and simulated quantity (i.e., the sum of squares).\footnote{In Figure 6, the initial point is omitted (plots start from iteration 1).}  The final squared 2-norm values are 70.94 $\si{rad}^2/\si{s}^2$ and 255.15 $\si{MW}^2$ for the rotational speed and active power, respectively.  Convergence can be visually seen in Figure~\ref{fig:conv}.  However, as the problem is nonlinear, convergence to a global optimum is not guaranteed~\cite{conv-opt}.  In this case, the degree of convergence can only be inferred by checking the resulting optimized parameters against the true parameter set $\tilde{p}$.  It is further noted that the average relative error achieved by the estimators can confirm the validity of the unbiased estimator hypothesis: if a sufficiently low average relative error is achieved across the iterations $N$, it can be said that the unbiased estimator assumption is adequate.

\subsubsection{Results}

\begin{table*}[ht]
  \centering
  \caption{Results of the Multi-Parameter Estimator Study.}
  \resizebox{\columnwidth}{!}{%
  \begin{tabular}{c|c|c|c|c|c|c|c|c|c|c}
    \toprule
    \multicolumn{3}{c}{Parameters}  
    & \multicolumn{4}{|c|}{Case 1: Frequency} & \multicolumn{4}{c}{Case 2: Active Power} \\ \cmidrule{1-11}
    Symbol & $\tilde{p}$ & $p_0$ & $\alpha$ & $\bar{p}$ & \begin{tabular}[c]{@{}c@{}}Avg.\\Relative \\ Error (\%) \end{tabular} & $\sigma^2_p$ [p.u.] & $\alpha$ & $\bar{p}$ & \begin{tabular}[c]{@{}c@{}}Avg.\\Relative \\ Error (\%) \end{tabular} & $\sigma^2_p$ [p.u.]  \\ \midrule
    $K$      & 20   & 18.5881  & 0.14    & 19.995 & 0.0266  & 2.26E-5 & 0.16 & 19.998 &  0.00862 & 2.09E-4 \\ \midrule
    $\sigma$ & 0.04 & 0.0376   & 0.44    & 0.0409 & 2.15    & 2.30E-7 & 0.0225 & 0.0407 & 1.65 & 1.33E-7  \\ \midrule
    $\delta$ & 0.8  & 0.8561   & 0.0275  & 0.8166 & 2.08    & 8.75E-5 & 0.0175 & 0.8136 & 1.70 & 5.61E-5 \\ \midrule
    $K_t$    & 1.5  & 1.5287   & 0.0225  & 1.531 & 2.08    & 3.07E-4 & 0.0175 & 1.525 & 1.64 & 1.85E-4  \\ \midrule
    $T_d$    & 2.4  & 2.6232   & 0.04    & 2.407 & 0.272   & 3.36E-2 & 0.0375 & 2.404 & 0.172 & 1.35E-2  \\ \midrule
    \multicolumn{3}{c}{Average $V_e$} & \multicolumn{4}{|c|}{4.78E-8} & \multicolumn{4}{c}{8.74E-10} \\ \midrule
    \multicolumn{3}{c}{Average $\sigma_p^2$} & \multicolumn{4}{|c|}{6.80E-3} & \multicolumn{4}{c}{2.80E-3} \\ \bottomrule
  \end{tabular}
  }
  \label{tab:multi-param-res}
\end{table*}

The results of the multi-parameter estimator study are shown in Table~\ref{tab:multi-param-res}.  Similarly to the previously shown case study, the confidence ellipsoid volume is computed after the parameter optimization run to maximize the accuracy of the nFIM estimate.  As the units of $T_d$ (and thus, the units of $\sigma_p$ for parameter $T_d$) are in seconds, the raw variance is normalized by the nominal time constant of $0.02 \textrm{~}\si{s}$ to produce the standard deviation of the parameter in per-unit values.

It is observed that the active power transient demonstrates both the lower confidence ellipsoid volume and the lower average parameter variances.  Therefore, we assert that the proposed Algorithm~\ref{alg:1} also shows coherency for the case of multi-parameter estimators.  It is observed that the average relative error is also lower for the active power measurements for all parameters; however, according to the theory, it is expected that this observation does not necessarily hold in all cases.  In practice, this finding implies that between two measurement sets of frequency and active power taken at the terminals of $SM_1$, the active power measurements are more suitable for identification of this parameter set, as their use would provide higher confidence in the obtained parameters when identified using a least-squares algorithm.

\section{Conclusions}\label{sec:conc}

In this paper, we propose the application of Fisher information theory to the identification of parameters in dynamic models of power systems.  As for the presented field of application, the constraints of Problem~\ref{eq:generic_id} are highly nonlinear, we derive a numerical approximation of the Fisher Information Matrix (nFIM) to be used for the selection of measurements.  This numerical approximation enables the application of FIM without the need for explicitly defining the log-likelihood function (see Eq.~\ref{eq:fim}) for power system transients, which further enables the use of commercially available time-domain numerical simulation software. We present that the measurements with the smallest confidence ellipsoid volume spanned by nFIM should be deployed to identify dynamic model parameters, and we show how the parameter perturbation $\alpha_k$ should be selected to maximize the accuracy of the nFIM computation. Although we apply the algorithm to governor and AVR parameters in this work, the method can also be used for identification of other parameters.

Case studies of single- and multi-parameter estimators using an IEEE 9-bus dynamic model demonstrate the coherency of the method with artificial measurements. Generally speaking, it is shown that estimators which exhibit a smaller confidence ellipsoid spanned by the numerical FIM exhibit smaller parameter variances.  For example, the single-parameter estimator for the turbine governor steady-state droop $\sigma$ exhibits the lowest confidence ellipsoid volume of the 5 parameters tested with 2.1E-8, and the lowest per-unit parameter variance with 100 experiments at 7.84E-11.  This parameter variance is shown to be higher than the numerical Cramer-Rao Lower Bound (nCRLB) of 1.11E-16.  For the case of multi-parameter estimators, it is shown that between the frequency and active power measurements, the active power transient should be used for model validation as it shows a lower confidence ellipsoid volume of 8.74E-10 and a lower average parameter variance of 2.8E-3 with 100 numerical experiments. In further work, this algorithm could be tested on real systems where the true parameters $\tilde{p}$ are known.

\bibliographystyle{IEEEtran}
\bibliography{paper}

\end{document}